\documentclass[fleqn,usenatbib]{mnras}

\usepackage{newtxtext,newtxmath}

\usepackage[T1]{fontenc}

\DeclareRobustCommand{\VAN}[3]{#2}
\let\VANthebibliography\thebibliography
\def\thebibliography{\DeclareRobustCommand{\VAN}[3]{##3}\VANthebibliography}

\usepackage{graphicx}	
\usepackage{amsmath}	

\title[A public code for astrometric microlensing]{A public code for astrometric microlensing with contour integration}
\author[Bozza, Khalouei and Bachelet]{
Valerio Bozza,$^{1,2}$\thanks{E-mail: valboz@sa.infn.it}
Elahe Khalouei $^{3}$, Etienne Bachelet $^{4}$ \\
$^{1}$Dipartimento di Fisica "E.R. Caianiello", Universit$\grave{a}$ di Salerno, Via Giovanni Paolo  132, Fisciano, I-84084, Italy\\$^{2}$Istituto Nazionale di Fisica Nucleare, Sezione di Napoli, Via Cintia, Napoli, I-80126, Italy\\
$^{3}$Department~of~Physics,~Sharif~University~of~Technology, P. O. Box 11365-9161, Tehran, Iran\\
$^{4}$Las Cumbres Observatory Global Telescope Network, 6740 Cortona Drive, suite 102, Goleta, CA 93117, USA.}

\begin{document}
\label{firstpage}
\pagerange{\pageref{firstpage}--\pageref{lastpage}}
\maketitle	
\begin{abstract}	
We present the first public code for the calculation of the astrometric centroid shift occurring during microlensing events. The computation is based on the contour integration scheme and covers single and binary lensing of finite sources with arbitrary limb darkening profiles. This allows for general detailed investigations of the impact of finite source size in astrometric binary microlensing. The new code is embedded in version 3.0 of \texttt{VBBinaryLensing}, which offers a powerful computational tool for extensive studies of microlensing data from current surveys and future space missions.
\end{abstract}
 

\section{Introduction}\label{intro}

Microlensing is a particular case of gravitational lensing occurring when the angular resolution is insufficient to detect any details of the images of a distant source other than an overall flux increase \citep{Paczynski1986}. However, even without a full resolution of the different images created by gravitational lensing, the centroid of the lensed source may move enough to allow a possible astrometric detection of microlensing \citep{Hog1995,Miyamoto1995,Walker1995}. For a single lens, the motion of the centroid describes an ellipse that can be used for an independent estimate of the angular Einstein radius of the lens, thus providing information on its mass and distance. Following these first studies, \citet{MaoWitt1998} studied the centroid motion caused by a single lens including an analytic treatment of the finite size of the source, showing that the centroid shift is reduced when finite source effects become important (see also \citet{Lee2010}). \citet{Safizadeh1999} considered the case in which the lens-star is accompanied by a planet. They calculated astrometric maps in analogy to magnification maps obtained by ray-shooting methods \citep{Wambsganss1997}: short bumps perturbing the elliptic trajectory of the centroid are typical signatures of the presence of planets. A more detailed numerical investigation was led by \citet{HanLee2002}, while \citet{Asada2002} proposed an analytic perturbative approach. For stellar binary lenses, instead, sudden jumps in the centroid occur when pairs of images are created or destroyed as the source crosses a caustic \citep{HanChunChang1999,Han2001}.

The actual detection of the centroid shift requires an astrometric accuracy better than the Einstein radius, which is less than 1 mas for observations toward the Galactic bulge. \citet{BodenShao1998} discussed the detectability of the centroid motion within the microlensing campaigns of their time, while \citet{DominikSahu2000} and \citet{BelokurovEvans2002} already envisioned the possibility of numerous detections with the Gaia mission\footnote{\url{https://www.esa.int/Science_Exploration/Space_Science/Gaia}}, followed by many other detailed predictions ranging between hundreds and thousands of expected detections \citep{Proft2011,Bramich2018,Rybicki2018,McGill2019,
McGill2020,Kuter2020}.  The astrometric signal has been searched in selected black hole candidates with the Keck telescopes \citep{Lu2016} and with HST \citep{Kains2017}. The incoming {\it Roman} telescope\footnote{\url{https://roman.gsfc.nasa.gov/}} will also detect astrometric microlensing for a large fraction of nearby lenses \citep{GouldYee2014}. 

The first firm detection of an astrometric microlensing signal came from observations from the HST on a nearby white dwarf that lensed a background star \citep{Sahu2017}. Soon after, Proxima Centauri was observed to give an astrometric microlensing signal on another background star with SPHERE at VLT \citep{Zurlo2018}. Finally, the new interferometer GRAVITY, mounted at VLT, has been able to separate the images in a microlensing event \citep{Dong2019}. 

Indeed, microlensing is becoming much more than just photometry and future facilities will be more and more sensitive to the astrometric signal. We can easily imagine that there will be frequent need of simultaneous modeling of the photometric and astrometric signals, which can only be done by codes that are able to calculate both photometry and astrometry at high speed. In the last few years, \texttt{VBBinaryLensing}\footnote{\url{http://www.fisica.unisa.it/GravitationAstrophysics/VBBinaryLensing.htm}} \footnote{\url{https://github.com/valboz/VBBinaryLensing}} stood out as the fastest public code for the calculation of microlensing effects \citep{Bozza1,Bozza2}. It is based on the contour integration principle \citep{Schramm1987,Dominik1995,Dominik1998,Gould1997}, which amounts to a conversion of the surface integral over the images into a line integral along the image boundaries according to Green's theorem\footnote{We remind that Green's theorem precedes Stokes' theorem, of which it represents a special case.}. This code is also at the computational core of all publicly available microlensing modeling platforms: \texttt{pyLIMA}\footnote{\url{https://github.com/ebachelet/pyLIMA}} \citep{Bachelet2017}, \texttt{muLAN}\footnote{\url{https://github.com/muLAn-project/muLAn}}, \texttt{MulensModel}\footnote{\url{https://github.com/rpoleski/MulensModel}} \citep{Poleski2019}. It is also used for real-time microlensing modeling in \texttt{RTModel}\footnote{\url{http://www.fisica.unisa.it/GravitationAstrophysics/RTModel.htm}}.

In this paper, we present a full implementation of the calculation of the astrometric shift of the centroid of the images within the contour integration framework of \texttt{VBBinaryLensing}. Section 2 recalls the basic formulae for calculating the microlensing magnification in the contour integration framework. Section 3 presents the same calculation for the astrometric centroid shift in binary microlensing. Section 4 deals with the point-lens microlensing, for which we adopt pre-calculated tables, followed by the conclusions. We also add a short appendix discussing our parameterization for full Keplerian orbital motion of the lens.

\section{Contour integration concept}

Let us consider a source at distance $D_S$ behind a gravitational lens at distance $D_L$ along the line of sight. Throughout the paper, all angular coordinates and angular distances will be expressed in units of the angular Einstein radius
\begin{equation}
\theta_E=\sqrt{\frac{4GM}{c^2}\left(\frac{1}{D_L}-\frac{1}{D_S}\right)} \label{thetaE}
\end{equation}
where $M$ is the total mass of the lens system, $G$ is Newton's constant, $c$ is the vacuum speed of light.

Let us indicate the angular coordinates of the source in the sky by $\mathbf{y}$. The gravitational lens will bend the light rays from the source, so that the observer will see the light coming from a different direction $\mathbf{x}$. In general, the original source position and the actual image position are related by a lens equation in the form

\begin{equation}
\mathbf{y}=\mathbf{f}(\mathbf{x}),
\end{equation}
where the function $\mathbf{f}$ encodes all the properties of the gravitational lens, including its mass distribution and geometrical configuration.

In order to find the positions of the images $\mathbf{x}$ for a given source $\mathbf{y}$, we need to solve the lens equation, which typically requires numerical techniques, except for some trivial cases.

If we have extended sources, we may try to ``shoot rays back'' from the observer to the source: by evaluating $\mathbf{f}(\mathbf{x})$ on a sufficiently dense grid in $\mathbf{x}$,  images will be found around those lucky $\mathbf{x}$ for which $\mathbf{f}(\mathbf{x})$ lands within the source boundary \citep{Kayser1986,Bennett1996,Wambsganss1997,Rattenbury2002,
Dong2006,Bennett2010}. Inverse-ray-shooting methods may thus calculate the area of the images and their centroid by knowledge of the positions of all individual image elements on the grid. This approach requires a large number of direct evaluations of the lens map $\mathbf{f}(\mathbf{x})$ on a 2-dimensional grid, with each evaluation being relatively fast.

An alternative approach is to reconstruct the image boundaries first and then, exploiting Green's theorem, convert the surface integral to a line integral on the boundary. The image boundaries can be found as a contour plot on a grid \citep{Schramm1987,Dominik1995,Dominik1998,Dominik2007}, which similar to the ray-shooting approach only requires the evaluation of $\mathbf{f}(\mathbf{x})$, but not its inversion. Alternatively, image boundaries can be found by inversion of the lens map on the source boundary \citep{Gould1997}. Compared to the inverse-ray-shooting method, an adaptive grid needs fewer evaluations of the lens map, since the grid is refined only along the boundaries. The other option (inversion of the lens map on the boundary) requires an even smaller number of calculations, but each individual calculation is much slower, since the inversion of the lens map by appropriate numeric algorithms is required. Different methods may have advantages or disadvantages for particular situations and may be complementary in many respects. A full comparison between these codes is beyond the scope of this paper, but it is worth remarking these differences whenever they become relevant to assess the efficiency of the codes.

Following the second strategy described above, for a uniform-brightness circular source centered at $\mathbf{y}_S$ and angular radius $\rho_*$, we can introduce a simple parameterization of the boundary
\begin{equation}
\mathbf{y}(\theta)=\mathbf{y}_{S}+ \rho_{*}  \left ( \begin{array}{c}
\cos \theta\\
\sin \theta\\
\end{array}\right), \label{sourceboundary}
\end{equation}
with $0\leq\theta<2\pi$.

By inversion of the lens map, we will find a certain number $N$ of images of this boundary. These boundaries $\mathbf{x}_I(\theta)$ will be functions of the parameter $\theta$ and will be labeled by an index $I\in\{1,...,N\}$ identifying the individual images.

Green's theorem states that for any regular functions $L_1$ and $L_2$, we can convert the surface integral on the image $I$ into a line integral on a counter-clockwise oriented boundary $\partial I$
\begin{equation}
\iint\limits_I \left(\frac{\partial L_2}{\partial x_1}- \frac{\partial L_1}{\partial x_2} \right)dx_1dx_2 = \int\limits_{\partial I} \left(L_1dx_1+L_2dx_2 \right), \label{Green}
\end{equation}

The orientation of the source boundary in Eq. (\ref{sourceboundary}) is counterclockwise, but the orientation of the image boundary is reversed for negative parity images. Therefore, if we use Green's theorem, we must multiply the right hand side by the parity $p_I=\pm 1$\footnote{In the adaptive grid framework \citep{Dominik2007}, this is not necessary, but one must distinguish holes from image boundaries.}. 

For a uniform source, since the surface brightness is conserved by gravitational lensing, the observed flux is simply the original source flux multiplied by a magnification factor given by the ratio of the total area of the images with the original area of the source. The area of an image is obtained by requiring that $\left(\frac{\partial L_2}{\partial x_1}- \frac{\partial L_1}{\partial x_2} \right)=1$, which leaves some freedom in the choice of the functions $L_1$ and $L_2$. For example, with $L_2=\frac{1}{2}x_1$ and $L_1=-\frac{1}{2}x_2$ \citep{Dominik1998}, Eq. (\ref{Green}) becomes 
\begin{equation}
A_I=-\frac{1}{2}p_I\int\limits_{\partial I} \mathbf{x} \wedge d\mathbf{x},
\end{equation}
where the wedge product between two vectors in the observer sky is a short notation for the following combination leading to a pseudoscalar quantity
\begin{equation}
\mathbf{x} \wedge \mathbf{w}\equiv x_1w_2 -x_2 w_1.
\end{equation}

With $L_2=0$ and $L_1=-x_2$, we have
\begin{equation}
A_I=-p_I\int\limits_{\partial I} x_{2} dx_{1}. \label{AI}
\end{equation}

The magnification for our uniform source of radius $\rho_*$ is then
\begin{equation}
\mu=\frac{1}{\pi \rho_*^2} \sum\limits_I A_I. \label{mu}
\end{equation}

In practice, we can sample the source boundary only on a finite set of points at $0=\theta_1< ... < \theta_n=2\pi$. For each $\theta_i$, we must invert the lens equation at the source boundary point $\mathbf{y}(\theta_i)$ and find the corresponding image boundary points $\mathbf{x}_{I,i}$ \citep{Gould1997}. The integral (\ref{AI}) is then approximated by
\begin{equation}
A_I=-\frac{p_I}{2}\sum\limits_i \left(x_{I,i+1,2}+ x_{I,i,2} \right)\left(x_{I,i+1,1} - x_{I,i,1} \right),
\end{equation}
which recalls the simple trapezium approximation of the Riemann integral \citep{Dominik1998,Gould1997}.

\citet{Bozza1} proposed to add a parabolic correction that catches the next-to-leading order contribution and leaves a residual of fifth order in $|d\mathbf{x}|$
\begin{equation}
dA_I^{(p)} = \frac{p_I}{24}\sum\limits_i \left[\left.\left(\mathbf{x}'_I\wedge \mathbf{x}''_I\right)\right|_{\theta_i} +\left.\left(\mathbf{x}'_I\wedge \mathbf{x}''_I\right)\right|_{\theta_{i+1}}\right]\Delta\theta_i^3, \label{dAIp}
\end{equation}
where $\Delta\theta_i=\theta_{i+1}-\theta_i$ and the primes denote derivatives of $\mathbf{x}_I(\theta)$ with respect to $\theta$.

\subsection{Contour integration in binary lensing}
\label{SecMag}

For binary lenses, we can adopt the complex notation for the source and image positions \citep{Witt1}
\begin{eqnarray}
&& \zeta=y_1+i y_2 \\
&& z=x_1+i x_2.
\end{eqnarray}

For a system of two point-lenses with separation $s$ and mass ratio $q$, the lens equation is
\begin{equation}
\zeta = z - \frac{1}{1+q}\left(\frac{1}{\bar z +
s}+\frac{q}{\bar z} \right). \label{LEQ complex}
\end{equation}

Here we have chosen the position of the minor lens as the origin of our coordinate frame. As also discussed by \citet{Bozza2}, this is the best choice for minimizing the numerical errors coming from cancellations of large terms in the lens equation.

For a given source position $\zeta$, the solutions of this equation give the corresponding images $z$. Using our parameterization for the source boundary (\ref{sourceboundary}), the quantities appearing in the parabolic correction can be calculated analytically in terms of derivatives of the lens equation
\begin{equation}
\mathbf{x}'\wedge \mathbf{x}'' = \left\{ \rho_*^2 + \mathrm{Im}\left[
(z')^2 \zeta' \frac{\partial^2 \bar \zeta}{\partial z^2} \right]
\right\} J^{-1}. \label{x'x''}
\end{equation}

\citet{Bozza1} also introduced error estimators to increase the sampling of the source boundary only where it is really needed to match the target accuracy. Finally, \citet{Bozza2} proposed an algorithm to decide whether the point-source approximation is acceptable or a full finite-source contour integration is necessary. This algorithm is based on the quadrupole approximation \citep{Pejcha2009} and the mutual separation of ghost images. Limb darkening of realistic stars is obtained by repeating the calculation on concentric circles centered on the source \citep{Gould1997}. The number and position of the circles is decided so as to match the target accuracy \citep{Bozza1}.  All these techniques are fully implemented in the current version of \texttt{VBBinaryLensing}, with the basic inversion of the lens equation realized by the algorithm by \citet{Skowron2012}. 

\section{Astrometry with contour integration}

High-resolutions observations have indeed proven to be able to measure the centroid shift during a microlensing event \citep{Sahu2017,Zurlo2018}. The centroid of the images for a uniform-brightness source can be defined as
\begin{equation}
\mathbf{\Theta}= \frac{\sum\limits_I \mathbf{X}_I}{\sum\limits_I A_I}, \label{Theta}
\end{equation}
where $A_I$ is just the area of the image $I$, as already defined in Eq. (\ref{AI}), which comes as a normalization factor. The numerator is the sum of the following integrals
\begin{equation}
\mathbf{X}_I=  \int\limits_{I} \mathbf{x} dx_1dx_2. \label{XIorig}
\end{equation}

Of course, the centroid shift is a vector with two components, which means that we have to calculate two more integrals with respect to the basic magnification calculation described in the previous section. However, even these two surface integrals can be converted to line integrals using Green's theorem. In fact, as shown by \citet{Dominik1998}, we just need to find new functions $L_1$ and $L_2$ such that $\left(\frac{\partial L_2}{\partial x_1}- \frac{\partial L_1}{\partial x_2} \right)=x_1$ for the first component and analogously for the second component of the centroid. These conditions leave some freedom in the choice of the two functions, which can be used so as to minimize the additional operations with respect to the magnification calculation. We finally choose $L_1=0, L_2=x_1^2/2$ to deal with the component $X_{I,1}$ and $L_1=-x_2^2/2, L_2=0$ for the component $X_{I,2}$. Following Green's theorem (\ref{Green}), Eq. (\ref{XIorig}), when split for the two components, becomes
\begin{eqnarray}
&& X_{I,1}= \frac{p_I}{2} \int\limits_{\partial I} x_1^2 dx_2 \label{XI1} \\
&& X_{I,2}= -\frac{p_I}{2} \int\limits_{\partial I} x_2^2 dx_1. \label{XI2}
\end{eqnarray}

In order to evaluate these two integrals, we use the same sampling of the source boundary described in Section \ref{SecMag}. Once we have the sampling of the image boundaries $\mathbf{x}_I$, we may just approximate the two integrals with sums of finite differences\footnote{By a different choice of the functions $L_1$ and $L_2$ it is possible to obtain more symmetric expressions \citep{Dominik1998}. However, our asymmetric choice halves the number of operations needed.}
\begin{eqnarray}
& X_{I,1}=\frac{p_I}{8} \sum\limits_i (x_{I,i+1,1}+x_{I,i,1})^{2}(x_{I,i+1,2}-x_{I,i,2}) &\\
& X_{I,2}=-\frac{p_I}{8} \sum\limits_i (x_{I,i+1,2}+x_{I,i,2})^{2}(x_{I,i+1,1}-x_{I,i,1}). &
\end{eqnarray}

Following the ideas laid out by \citet{Bozza1}, we may add a parabolic correction that pushes the accuracy of our approximation to $\Delta\theta^3$, similarly to the magnification calculation. 

\begin{eqnarray}
\begin{split}
dX_{I,1}^{p}=\frac{p_I}{24} \bigg[ \Big(x'^{2}_{I,1} x'_{I,2} +  x_{I,1} (\mathbf{x}'_{I} \wedge \mathbf{x}''_{I} ) \Big) |_{\theta_{i}} \\
+  \Big(x'^{2}_{I,1} x'_{I,2} +  x_{I,1} (\mathbf{x}'_{I} \wedge \mathbf{x}''_{I} ) \Big) |_{\theta_{i+1}} \bigg]\Delta\theta^{3}
\end{split}
\end{eqnarray}

\begin{eqnarray}
\begin{split}
dX_{I,2}^{p}=-\frac{p_I}{24} \bigg[ \Big(x'^{2}_{I,2} x'_{I,1} +  x_{I,2} (\mathbf{x}'_{I} \wedge \mathbf{x}''_{I} ) \Big) |_{\theta_{i}} \\
+  \Big(x'^{2}_{I,2} x'_{I,1} +  x_{I,2} (\mathbf{x}'_{I} \wedge \mathbf{x}''_{I} ) \Big) |_{\theta_{i+1}} \bigg]\Delta\theta^{3}
\end{split}
\end{eqnarray}

These corrections are obtained after expanding Eqs. (\ref{XI1})-(\ref{XI2}) in powers of $\Delta \theta$ to the third order. Note that the quantities appearing in these two corrections are already available from the parabolic correction to magnification (\ref{dAIp}). The implementation of accurate astrometry does not require too much additional computation, after all.

\subsection{Critical points}

Up to now we have only considered regular images leaving aside the creation and destruction of pairs of images. In case two images are destroyed in the interval $[\theta_i,\theta_{i+1}]$, we have some differences in the terms approximating our integrals. For the magnification, in the case of destruction of two images, we use \citep{Bozza1}
\begin{equation}
dA_c^{(t)} = \frac{1}{2} \left( x_{-,i,2} + x_{+,i,2}\right)
\left(x_{-,i,1} - x_{+,i,1} \right),
\end{equation}
for the first order term and
\begin{equation}
dA_c^{(p)} = \frac{1}{24}\left[\left(\mathbf{x}'_{+,i}\wedge \mathbf{x}''_{+,i}\right) -\left(\mathbf{x}'_{-,i}\wedge \mathbf{x}''_{-,i}\right)
\right] \widetilde{\Delta\theta}^3, \label{dAIpc}
\end{equation}
for the parabolic correction. In these expressions, $\mathbf{x}_+$ is the positive parity image and $\mathbf{x}_-$  is the negative parity image that participate in the destruction event. We also note that $\Delta \theta$ is replaced by
\begin{equation}
\widetilde{\Delta\theta} =  \frac{\left| \vec x_{+,i} - \vec
x_{-,i} \right| }{\sqrt {|\vec x'_{+,i} \cdot \vec x'_{-,i}|}},
\label{tildedtheta}
\end{equation}
since the two images are annihilated at some value $\theta_c$ between $\theta_i$ and $\theta_{i+1}$. This value is estimated using the distance of the last points of the two images and their current derivatives. Analogous expressions can be used for the creation of two images. 

As detailed by \citet{Bozza1}, error estimators can be constructed for intervals in which creation/destruction of images occur, which also work when the source intercepts the tip of a cusp. A more subtle case is a source nearly tangent to a fold, for which we check the mutual distance of the two spurious roots of the fifth order polynomial associated with the lens equation. In comparison, the adaptive grid approach used by \citet{Dominik2007} needs to introduce seeds in the grid for each image center. For partial images existing when only a small slice of the source is inside the caustic, it is necessary to add the points with minimum distance between the source center and the caustics as seeds.

For astrometry everything goes very similar. We use
\begin{eqnarray}
& dX_{c,1}^{(t)}=\frac{1}{8} (x_{+,i,1}+x_{-,i,1})^{2}(x_{+,i,2}-x_{-,i,2}) &\\
& dX_{c,2}^{(t)}=-\frac{1}{8} (x_{+,i,2}+x_{-,i,2})^{2}(x_{+,i,1}-x_{-,i,1}). &
\end{eqnarray}
for the first order terms and
\begin{eqnarray}
\begin{split}
dX_{c,1}^{p}=\frac{1}{24} \bigg[ \Big(x'^{2}_{+,i,1} x'_{+,i,2} +  x_{+,i,1} (\mathbf{x}'_{+,i} \wedge \mathbf{x}''_{+,i} ) \Big) \\
-  \Big(x'^{2}_{-,i,1} x'_{-,i,2} +  x_{-,i,1} (\mathbf{x}'_{-,i} \wedge \mathbf{x}''_{-,i} ) \Big)\bigg]\Delta\theta^{3}
\end{split}
\end{eqnarray}

\begin{eqnarray}
\begin{split}
dX_{c,2}^{p}=-\frac{1}{24} \bigg[ \Big(x'^{2}_{+,i,2} x'_{+,i,1} +  x_{+,i,2} (\mathbf{x}'_{+,i} \wedge \mathbf{x}''_{+,i} ) \Big)  \\
-  \Big(x'^{2}_{-,i,2} x'_{-,i,1} +  x_{-,i,2} (\mathbf{x}'_{-,i} \wedge \mathbf{x}''_{-,i} ) \Big) \bigg]\Delta\theta^{3},
\end{split}
\end{eqnarray}
for the parabolic corrections.

\begin{figure*}
	\begin{center}
		\includegraphics[angle=0,width=0.9\textwidth,clip=]{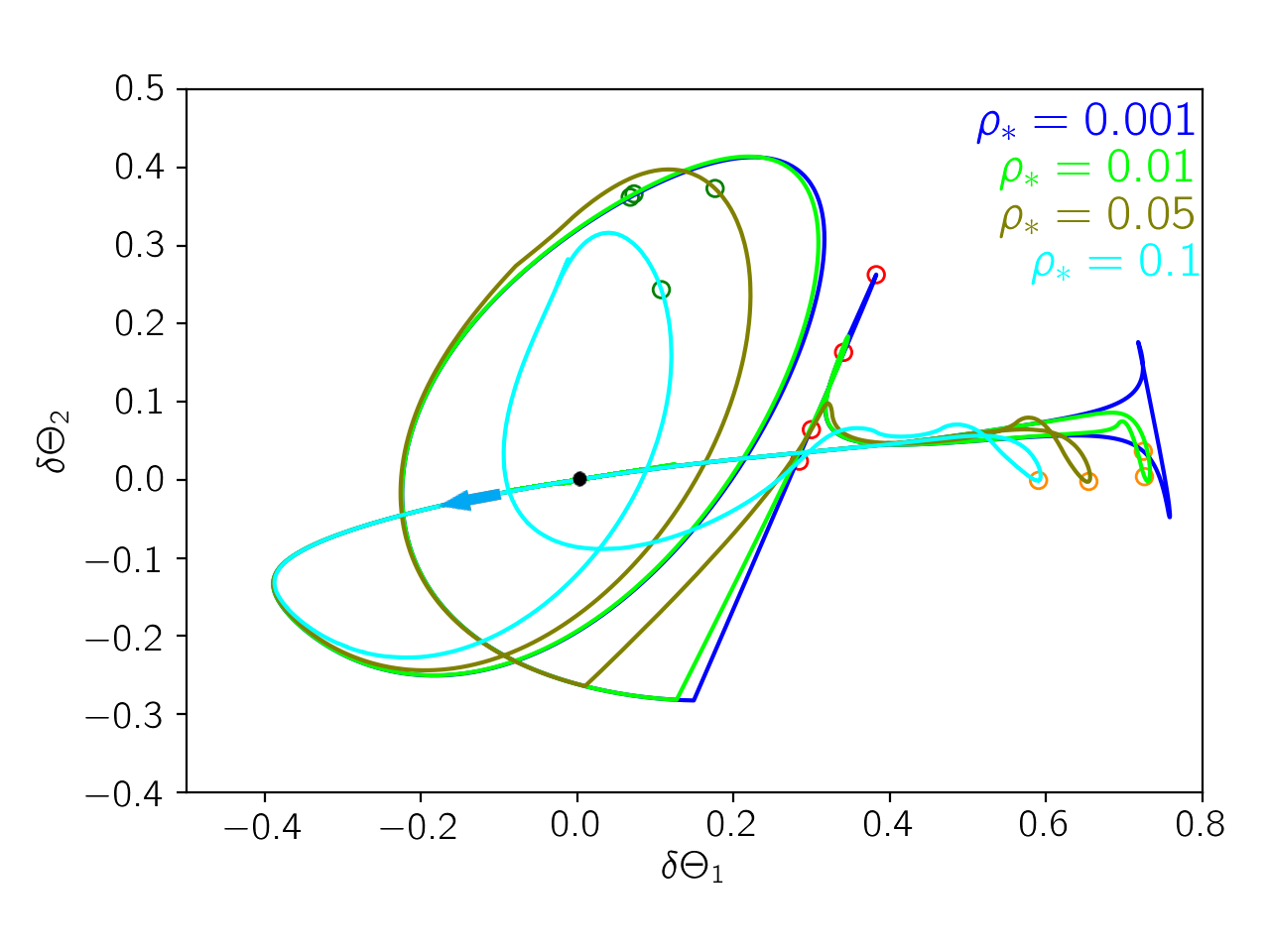}
		\caption{Centroid trajectories obtained with different source sizes $\rho_*$ in the geometry shown in Fig. \ref{figsource}. As specified in the text, all angular coordinates are in units of $\theta_E$. The circles of different colors mark the centroid positions corresponding to the source positions marked with the same color in Fig. \ref{figsource}.}
		\label{figcentroid}
	\end{center}
\end{figure*}

\begin{figure}
	\begin{center}
	\includegraphics[angle=0,width=\columnwidth,clip=]{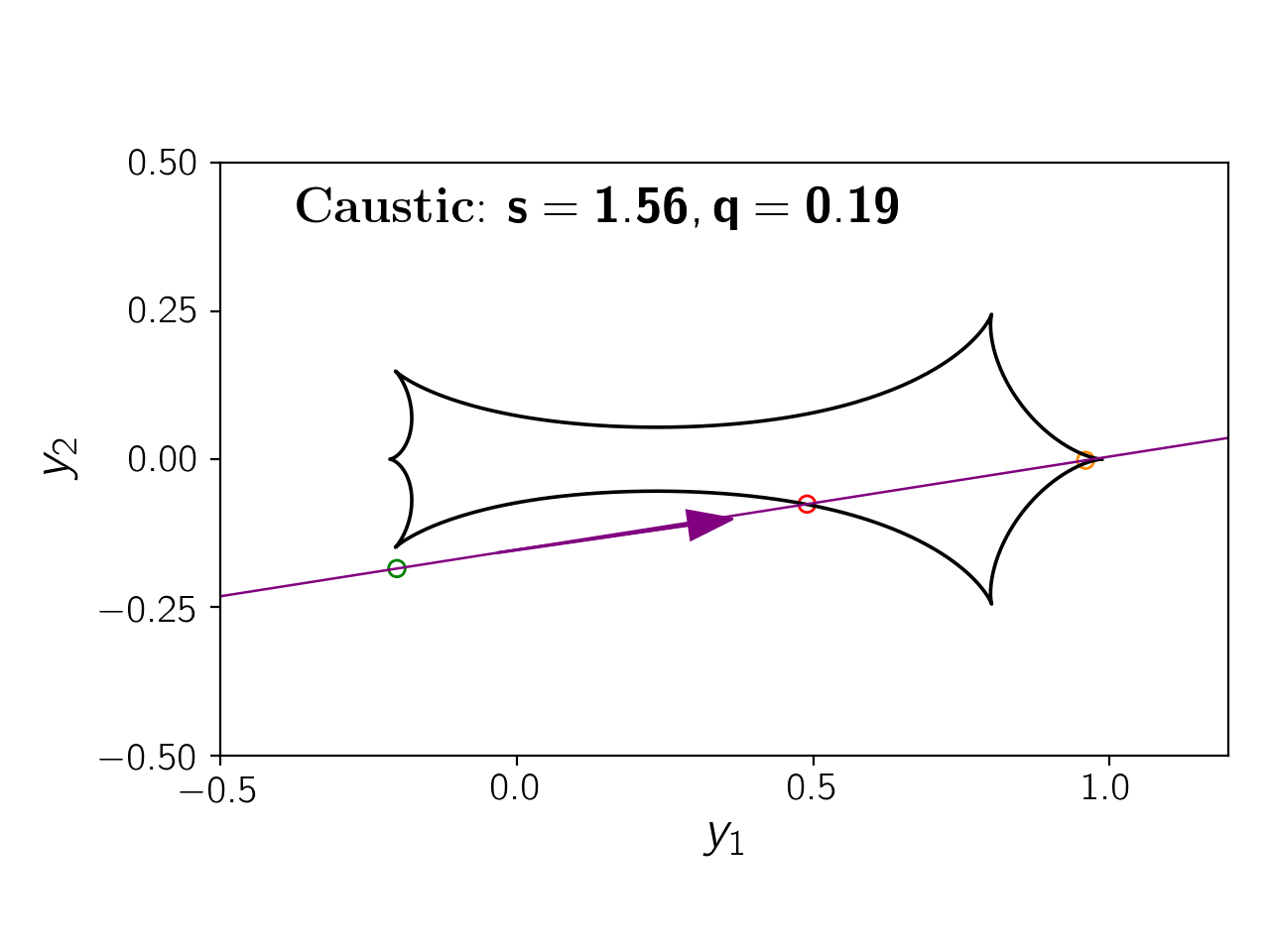}
		\caption{Source trajectory and caustic geometry used to generate Fig. \ref{figcentroid}. $s$ is the angular separation of the two lenses and $q$ is the mass ratio, with the lower-mass lens lying on the right. As specified in the text, all angular coordinates are in units of $\theta_E$. Circles mark specific positions useful to follow the centroid trajectories in Fig. \ref{figcentroid}.}
		\label{figsource}
	\end{center}
\end{figure}

\subsection{A first example: dependence of centroid motion on source size}

In Fig. \ref{figcentroid} we show a first example of use of our code for the calculation of the centroid trajectories corresponding to the source trajectory in Fig. \ref{figsource}, crossing a caustic in the intermediate regime. Note that $\mathbf{\Theta}$, as defined in Eq. (\ref{Theta}) is the centroid position relative to the lens. Actually, the observed blend is made up by the superposition of the lens and the microlensed source. If the lens makes a negligible contribution to the total flux, then what is actually observed is the centroid shift relative to the original source position. Therefore, in this and the following figures we plot $\mathbf{\delta \Theta} \equiv \mathbf{\Theta} -\mathbf{y}_S$. The generalization to a luminous lens is straightforward \citep{DominikSahu2000}.

The chosen example is particularly illuminating for our purposes. In order to explain the centroid shift, we better look at the trajectory corresponding to the smallest source $\rho_*=0.001$, shown in blue. When the source is still far from the lens, the centroid starts moving to the left, where the source comes from. Then, all tracks rotate clockwise downward. When the source approaches a cusp (see Fig. \ref{figsource}), the centroid makes a very large loop and comes back to the same region. This loop is due to the fact that the negative parity image generated by the left lens is highly magnified. Note that larger sources describe progressively smaller loops. 

Then the source enters the caustic through a fold and the centroid suddenly jumps to $(0.4,0.2)$. The jump is anticipated for larger sources, for $\rho_*=0.1$ it is even confused with the first cusp approach. 

As the source moves inside the caustic, the centroid moves to the right. When the source exits the caustic through a cusp, the $\rho_*=0.001$ centroid shows some structure at the extreme right, corresponding to a fold exit very close to the cusp, while larger sources progressively smooth these spikes and remain closer to the origin. As the source finally moves away, the centroid returns toward the origin.

For a point-like source, the centroid of the images would experience a discontinuous jump every time the source crosses a caustic \citep{HanChunChang1999,Han2001}. This happens because the newly formed images dominate the magnification and attract the centroid especially when they are close to the critical curve. For larger sources, this effect is smoothed progressively. In general, the overall centroid motion becomes less prominent and less informative for large sources.

\begin{figure}
	\begin{center}
	\includegraphics[angle=0,width=9cm,clip=]{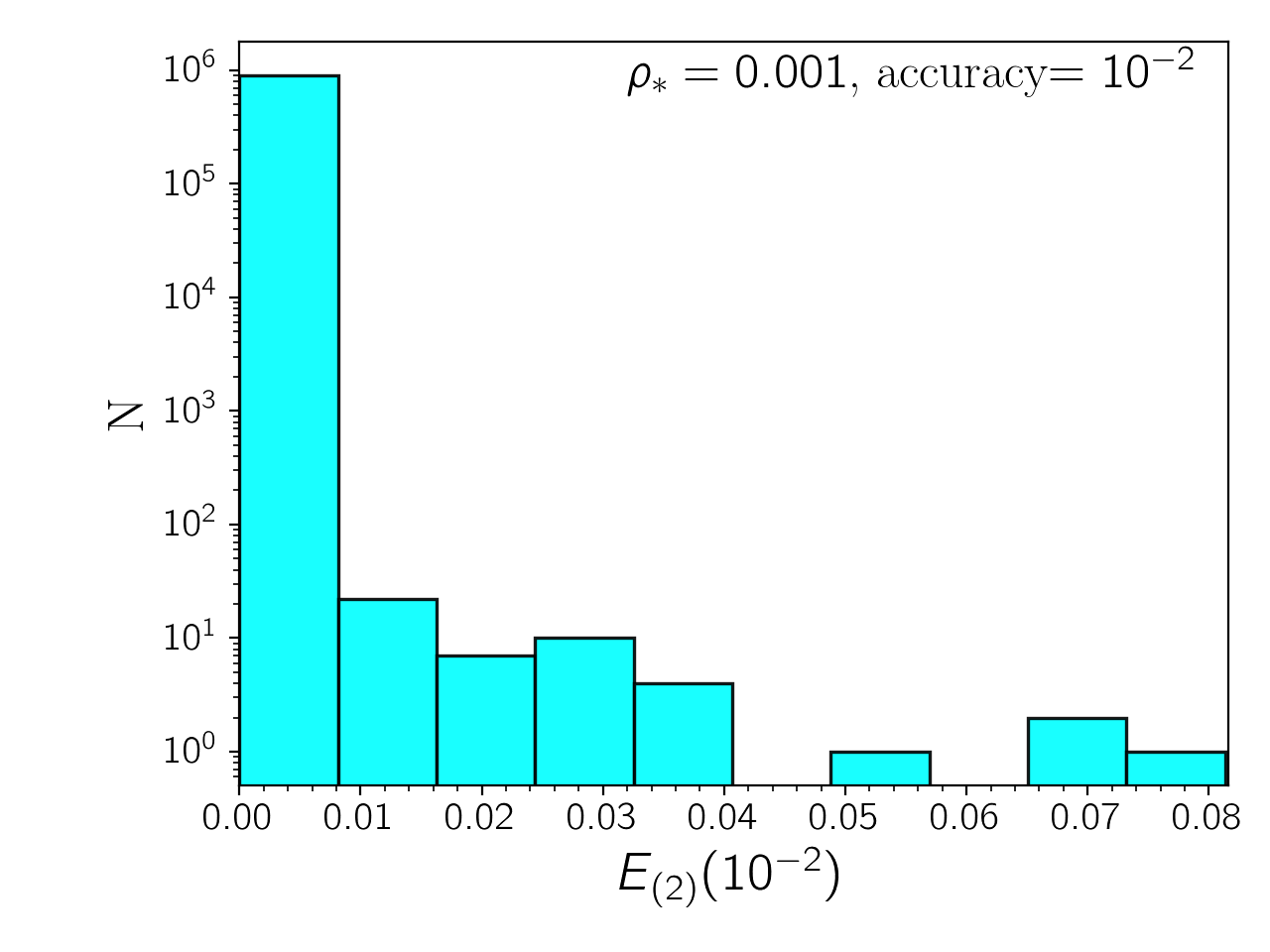}
	\includegraphics[angle=0,width=9cm,clip=]{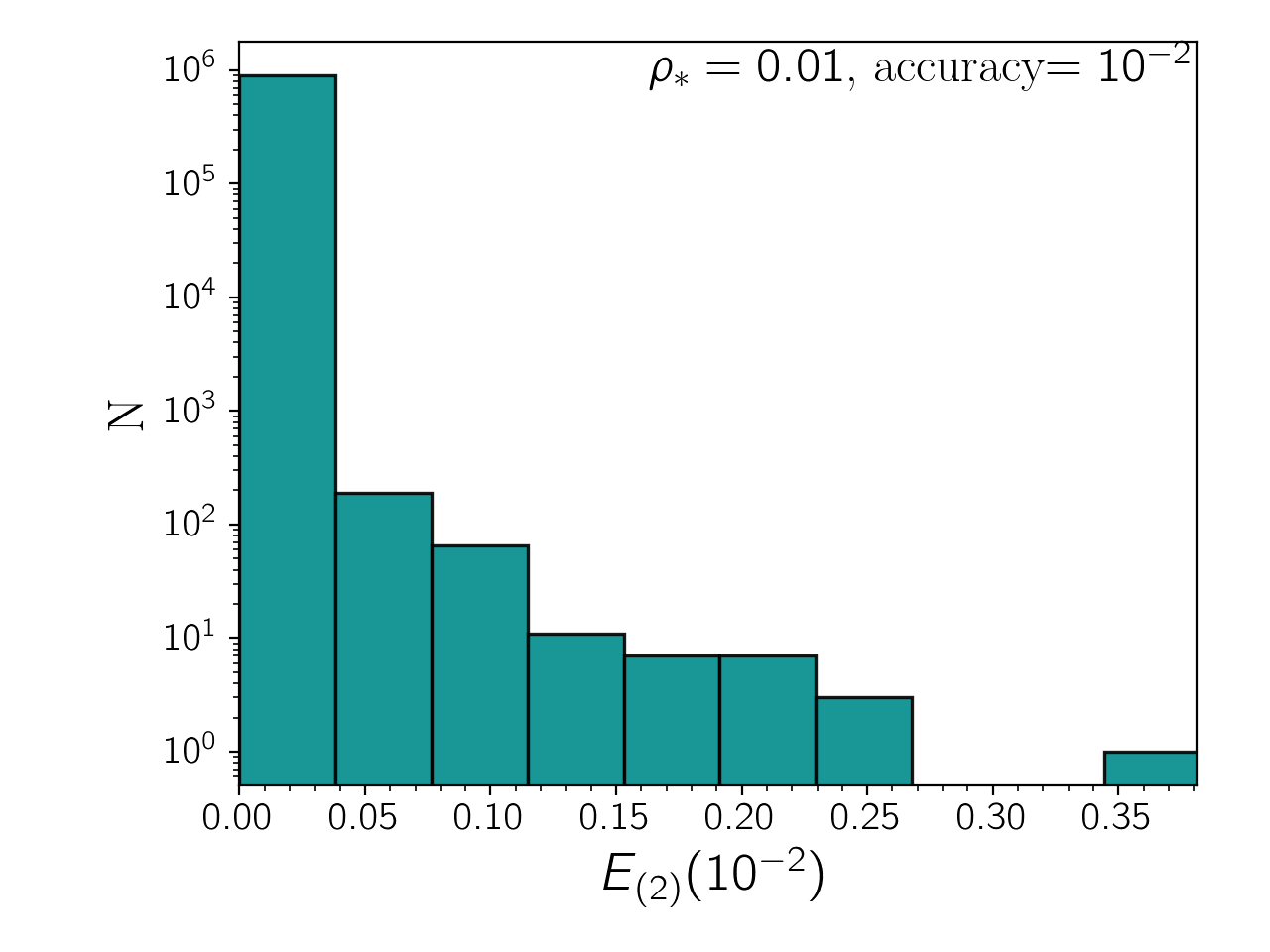}
	\includegraphics[angle=0,width=9cm,clip=]{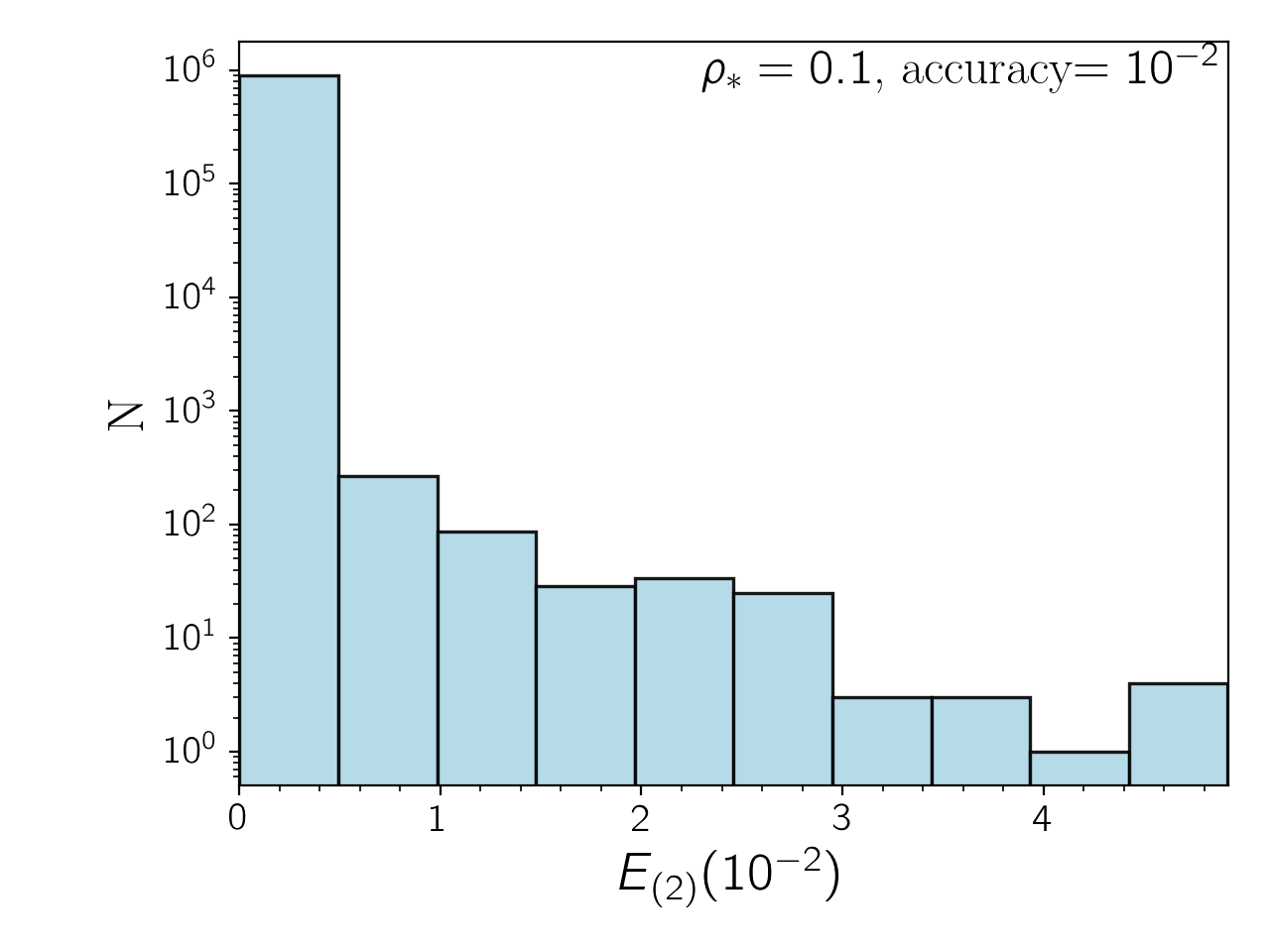}
		\caption{Distribution of the observed errors in the centroid shifts for a target accuracy $\delta=10^{-2}$ in magnification. $E_{(2)}$ is the error as defined in Eq. (\ref{centroiderror}). $N$ is the number of points in our grid falling in a given bin.}
		\label{figerrors2}
	\end{center}
\end{figure}

\begin{figure}
	\begin{center}
	\includegraphics[angle=0,width=9cm,clip=]{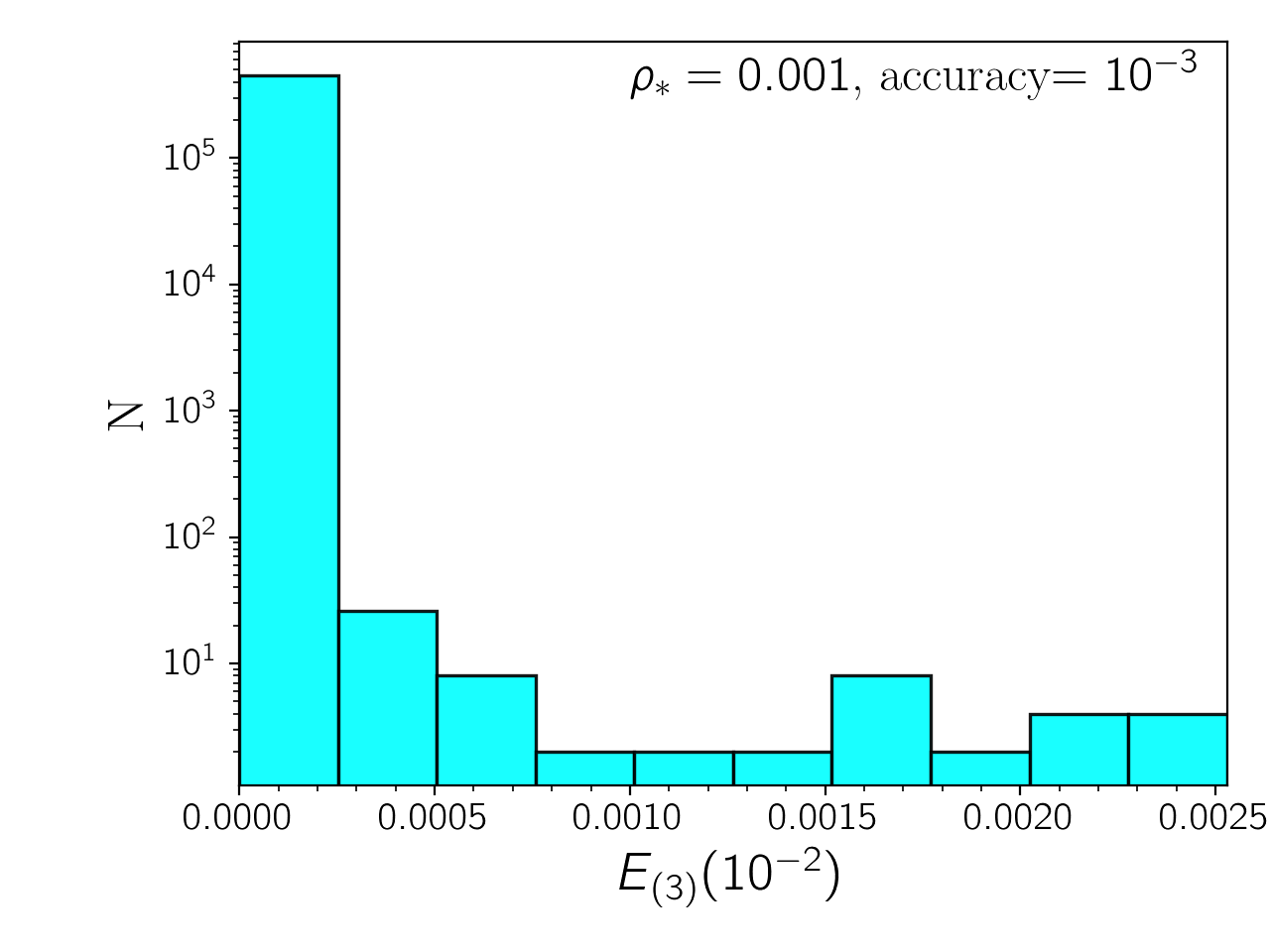}
	\includegraphics[angle=0,width=9cm,clip=]{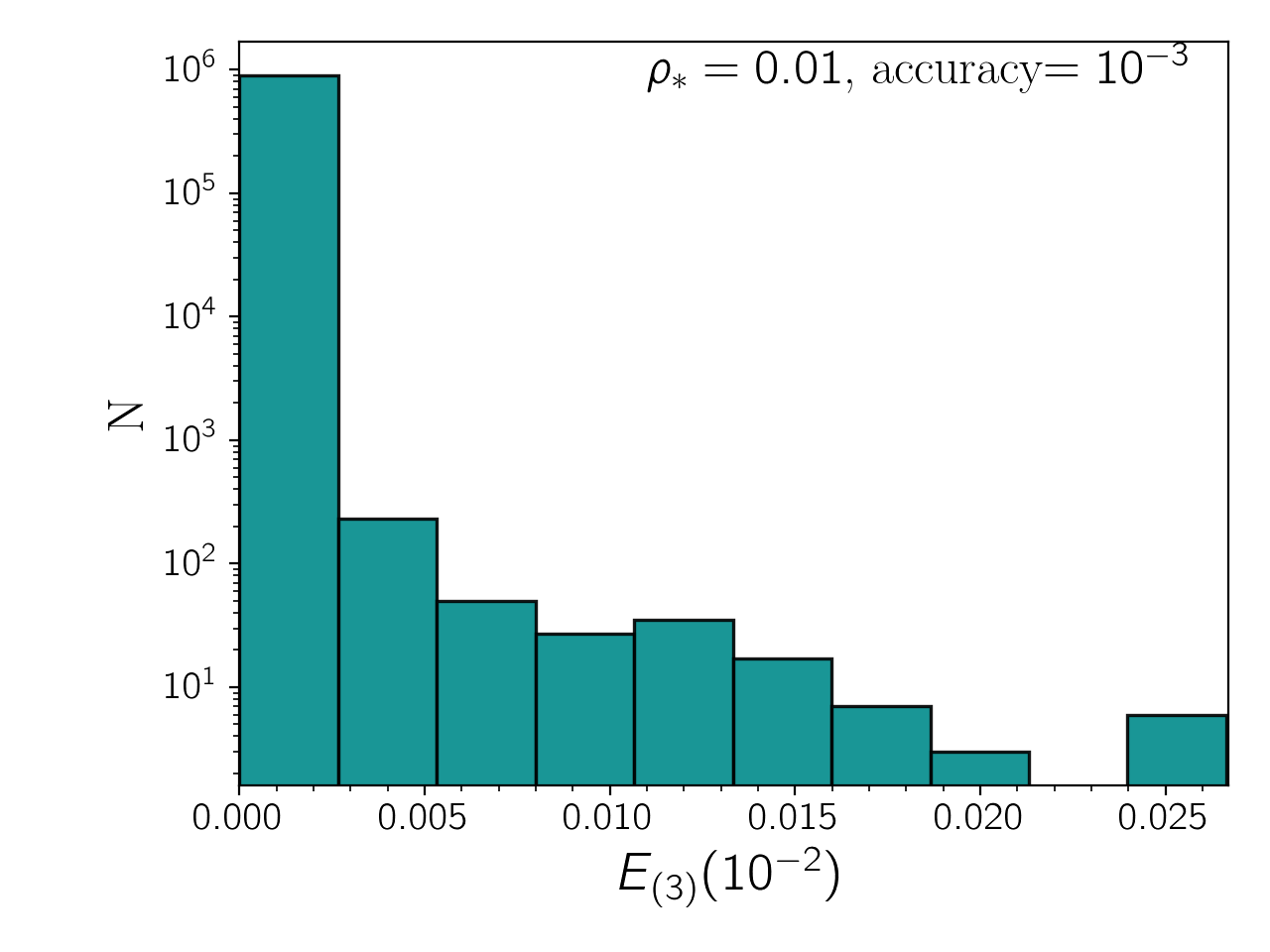}
	\includegraphics[angle=0,width=9cm,clip=]{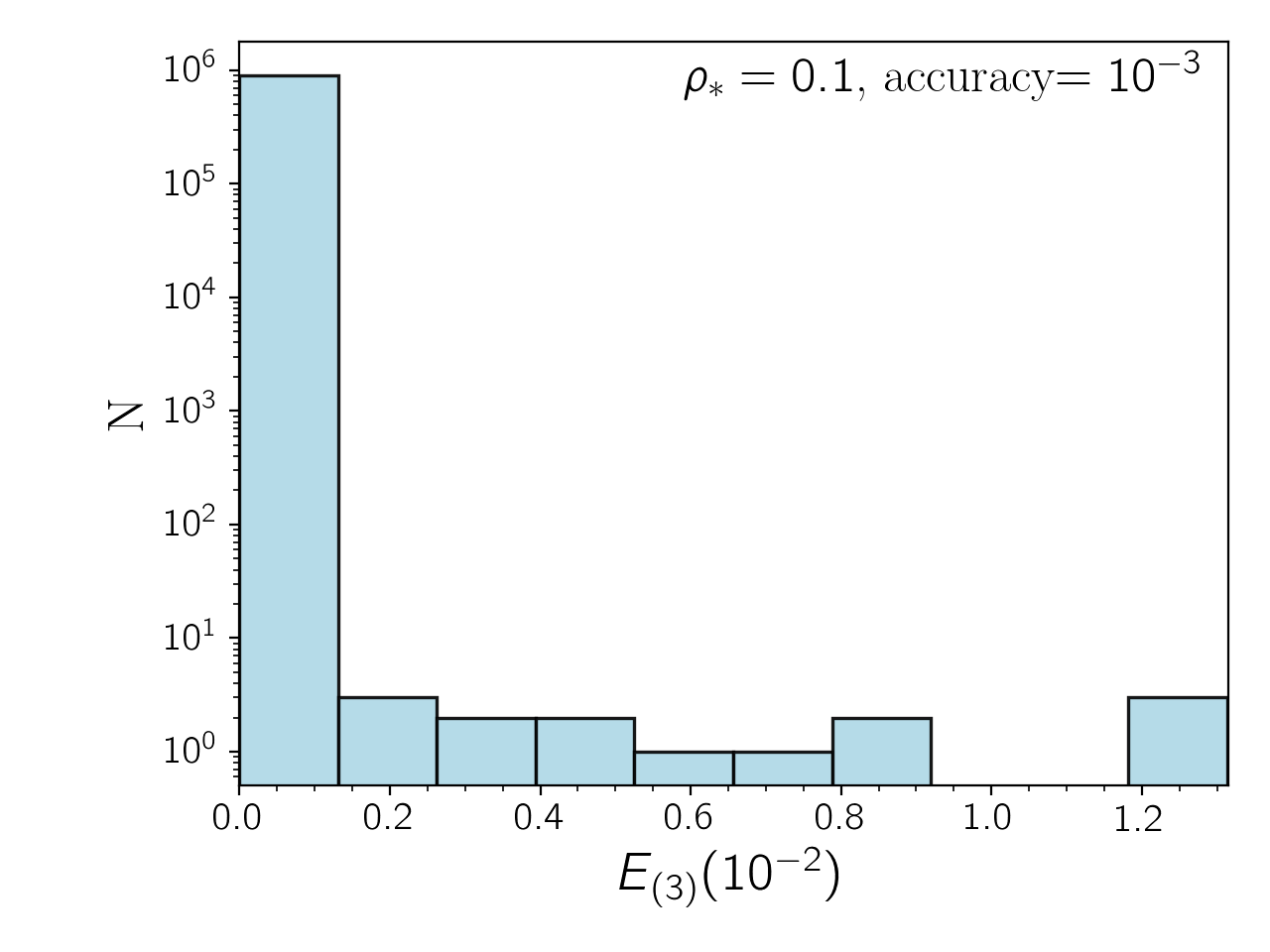}
		\caption{Distribution of the observed errors in the centroid shifts for a target accuracy $\delta=10^{-3}$ in magnification. $E_{(3)}$ is the error as defined in Eq. (\ref{centroiderror}). $N$ is the number of points in our grid falling in a given bin.}
		\label{figerrors3}
	\end{center}
\end{figure}

\subsection{Accuracy of the calculation} \label{Sec accuracy}

In \texttt{VBBinaryLensing} the accuracy is controlled by several error estimators on every individual arc of the source boundary \citep{Bozza1}. This allows to drive the sampling toward more problematic sections of the source and its corresponding images. The code stops the sampling when a target accuracy $\delta$ (specified by the user) is reached in the magnification.

In order to evaluate the centroid accuracy reached for a given target accuracy in magnification, we define $\mathbf{\delta \Theta}^{(i)}$ the centroid shift calculated at a target accuracy $\delta=10^{-i}$. For example, $\mathbf{\delta \Theta}^{(2)}$ is the centroid at $\delta=10^{-2}$. This is compared with the result obtained at a much higher accuracy, which is considered as ``exact'' (we use $\mathbf{\delta \Theta}^{(6)}$ for this purpose). The difference between the two calculations gives the observed error
\begin{equation}
E_{(i)}=\left|\mathbf{\delta \Theta}^{(i)}- \mathbf{\delta \Theta}^{(6)} \right|. \label{centroiderror}
\end{equation}

In order to illustrate the typical errors, we run the calculation on equally spaced grids of $10^5$ source positions on 10 configurations with $0.1\leq s \leq 2$ and $10^{-3} \leq q \leq 1$. The results are reported in the distributions shown in Fig. \ref{figerrors2} for $E_{(2)}$ and Fig. \ref{figerrors3} for error$E_{(3)}$.  Note that the frequencies are reported in log-scale. Then the vast majority of the points in the grid have a very low error compared to a tail that extends to larger errors. This behavior stems from the simple fact that most of the points in the grid fall far from the caustics, where the point-source approximation is already viable. The real test comes from the points in the tail, whose error must be taken under control.

Indeed, we see a clear linear scaling of the error distributions both with the target accuracy $\delta$ and with the source radius $\rho_*$. A reasonable formula that accounts for the observed accuracy in the centroid shifts is
\begin{equation}
\delta_{ast}=50 \rho_* \delta. \label{Astroacc}
\end{equation}

At this point we note that for typical photometric accuracy of $10^{-2}$ and $\rho_*=0.01$, the astrometric error remains below $5\times 10^{-3}$. For an angular Einstein radius of the order  mas, the reached accuracy is of the order $5 \mu$as, which is even  below Gaia accuracy. From this example, we realize that in general more stringent requirements will always come from photometry rather than astrometry. Therefore, we will not define error estimators for the individual contribution of each source boundary arc to the centroid error budget, as we did for photometric magnification, since this would make the algorithm complicated beyond any practical purposes. We prefer to set only one target accuracy for photometry and use Eq. (\ref{Astroacc}) to check that this is sufficient for astrometry as well. This formula is therefore very precious to guide the user in his/her choice for the target accuracy $\delta$. For example, if an astrometric accuracy of $\delta_{ast}=10^{-2}$ is needed and the source size is $\rho_*=0.01$, the target accuracy must be set to $\delta<\frac{1}{50}=0.02$.

 \begin{figure}
	\begin{center}
		\includegraphics[angle=0,width=0.49\textwidth,clip=]{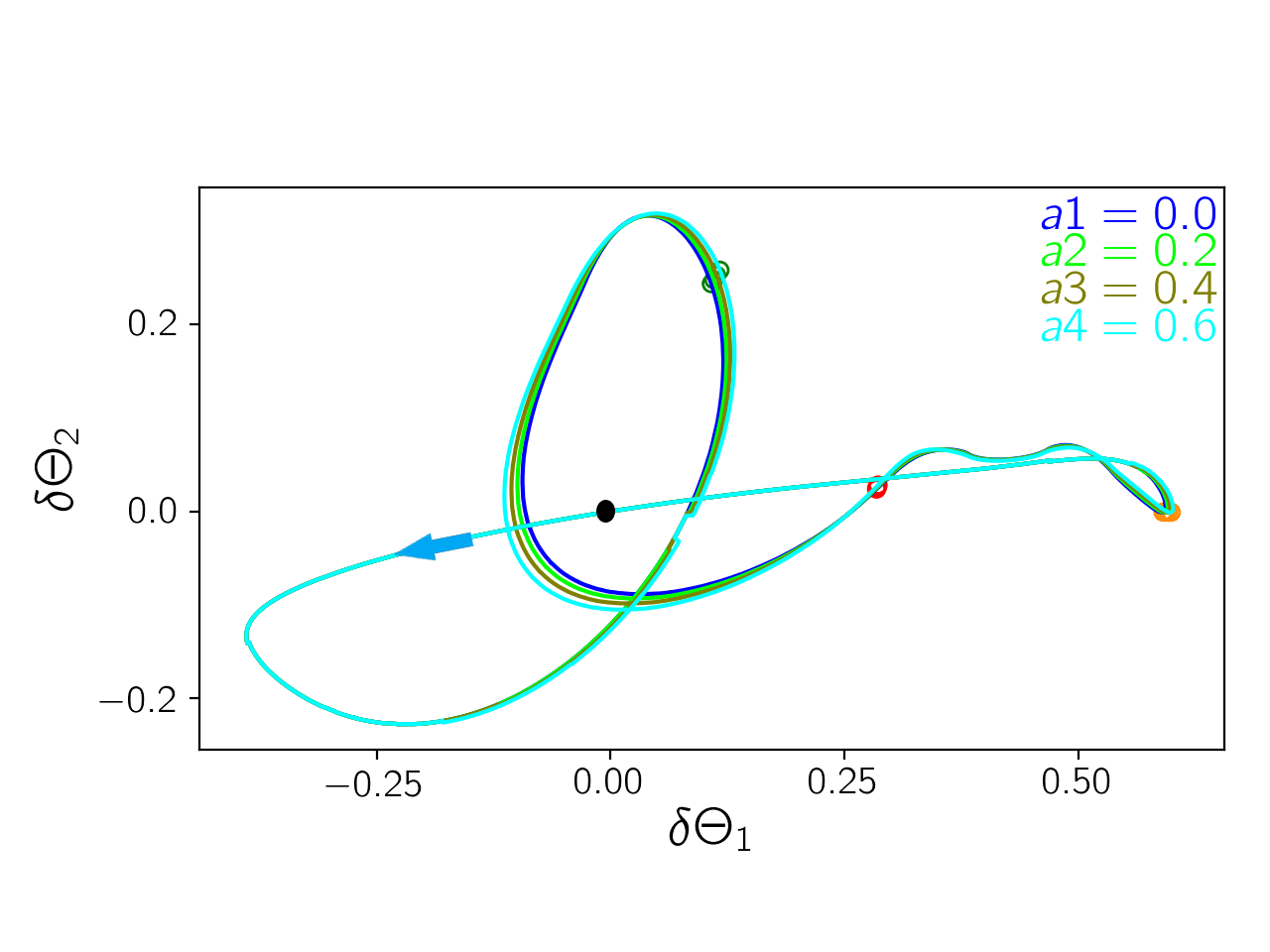}
		\caption{Centroid trajectories for the same geometry of Fig. \ref{figsource} with $\rho_*=0.5$ and several linear limb darkening coefficients.}
		\label{figlimb}
	\end{center}
\end{figure}

\begin{figure*}
	\begin{center}
		\includegraphics[angle=0,width=0.45\textwidth,clip=]{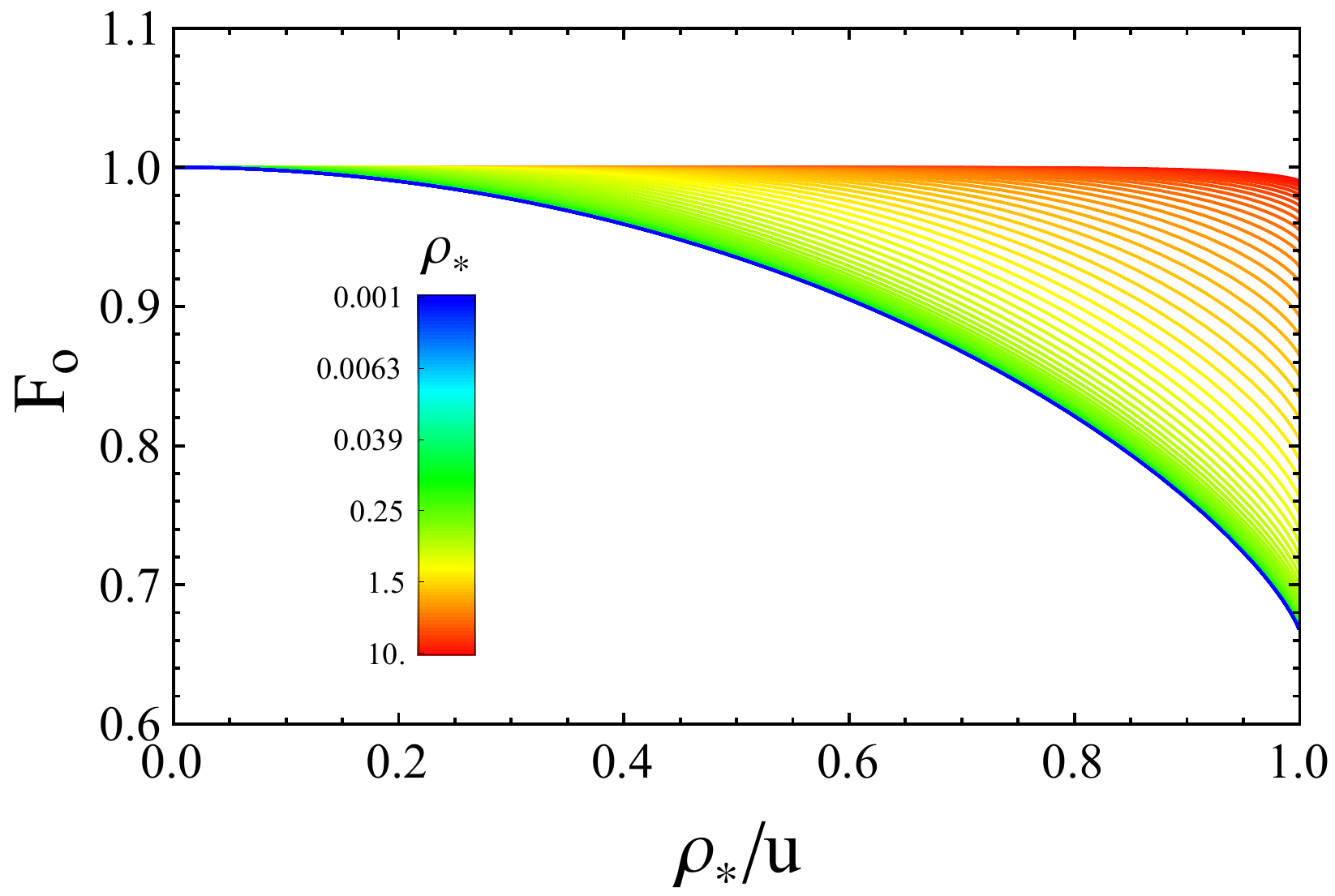}
		\includegraphics[angle=0,width=0.45\textwidth,clip=]{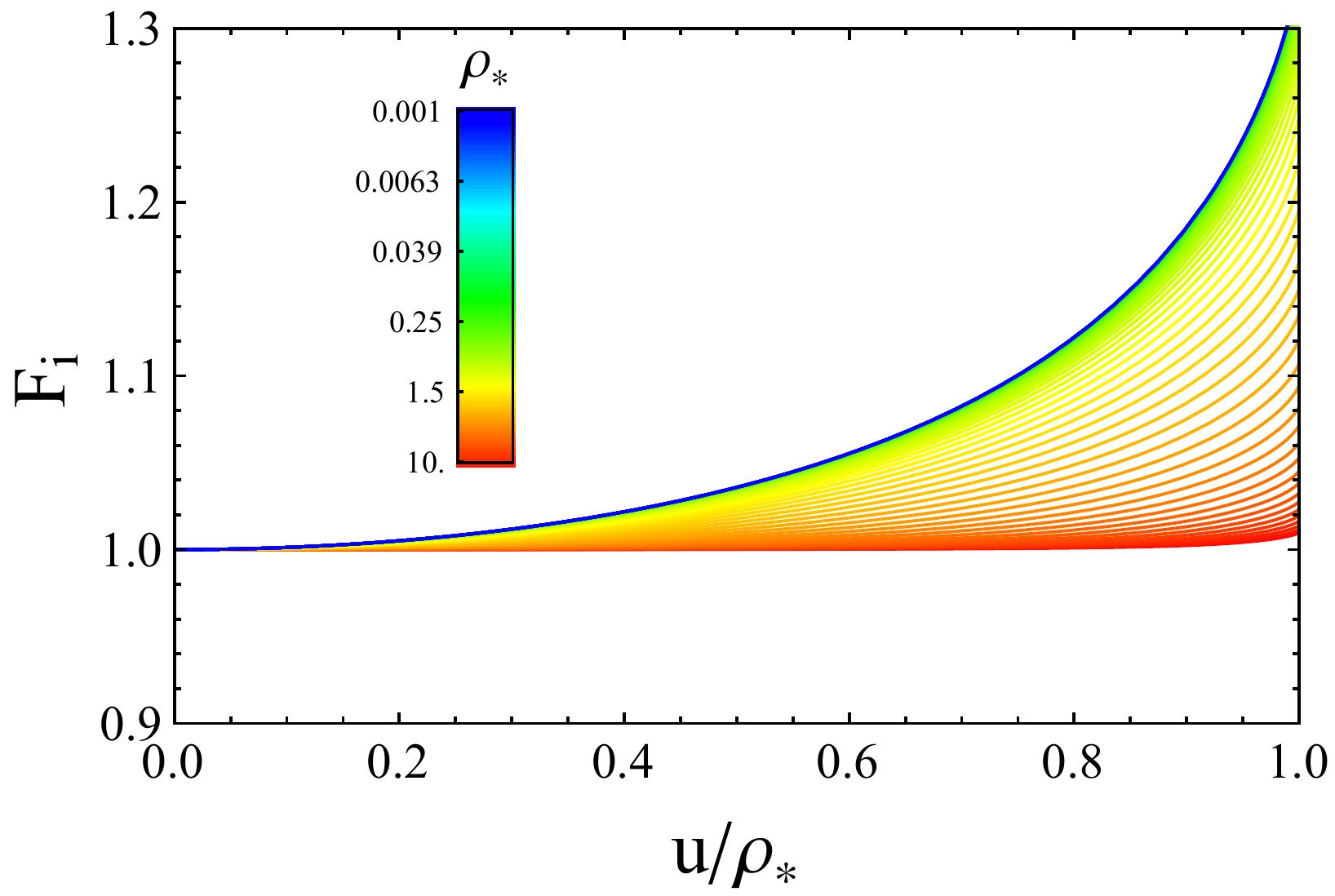}
		\caption{Finite-size corrections $F_{0}$ and $F_{i}$, as defined in Eq. (\ref{LDcorrection}) in terms of $\rho_{*}/u$ and $u/\rho_{*}$ respectively.}
		\label{figESPL}
	\end{center}
\end{figure*}
\subsection{Limb darkening} \label{Sec limb}

Up to now, we have only considered uniform-brightness sources with radius $\rho_*$. Actually, microlensed stars have a limb-darkened profile, which affects both the magnification and the centroid shift. Following \citet{Bozza1}, let us consider a generic intensity profile
\begin{equation}
I(\rho)= \bar{I}  f(\rho/\rho_{*}), 
\end{equation}
where $ \bar{I} $ is the average surface brightness. The profile function $f(r)$ is normalized to 1
\begin{equation}
\int\limits_0^1 2rf(r)dr = 1.
\end{equation}
Its cumulative function is
\begin{equation}
F(r)=2\int_{0}^{r} dr' r' f(r').
\end{equation}

The magnification for a limb-darkened extended source can be written as an integral of the point-source magnification, here denoted by $\mu_0$
\begin{equation}
 \mu_{LD} = \frac{1}{\pi}\int\limits_{0}^{1} r f(r)
 dr\int\limits_0^{2\pi}
 d\theta\mu_0(r \rho_*,\theta). \label{muLD}
\end{equation}
As a particular case, for a constant profile we simply recover the uniform-brightness magnification (\ref{mu}) in the form
\begin{equation}
 \mu = \frac{1}{\pi}\int\limits_{0}^{1} r
 dr\int\limits_0^{2\pi}
 d\theta\mu_0(r \rho_*,\theta). 
\end{equation}

%

\citet{Bozza1} showed that the limb-darkened magnification (\ref{muLD}) can be well-approximated  by a sum of uniform-brightness magnifications calculated on concentric annuli. If such annuli have normalized radii $0<r_1< ... r_n=1$, we set
\begin{equation}
\mu_{LD}\simeq \sum_{i} \tilde{M}_{i}
\end{equation}
where 
\begin{equation}
\tilde M_i= f_i\left[\mu_i r_{i}^2- \mu_{i-1} r_{i-1}^2\right]
\end{equation}
is the contribution of the $i$-th annulus with internal radius $r_{i-1}$ and external radius $r_i$. In this expression,
\begin{equation}
f_i=\frac{F(r_i)-F(r_{i-1})}{r_{i}^2- r_{i-1}^2 }. \label{fi}
\end{equation}
is the average value of the profile in the annulus and $\mu_i$ is the uniform-brightness magnification of the disk with radius $r_i$.

In complete analogy with Eq. (\ref{muLD}), the centroid of the images of the limb-darkened source can be expressed as \citep{MaoWitt1998}
\begin{equation}
 \mathbf{\Theta}_{LD} = \frac{1}{\pi}\int\limits_{0}^{1} r f(r)
 dr\int\limits_0^{2\pi}
 d\theta\mu_0(r \rho_*,\theta) \mathbf{\Theta}_0 (r \rho_*,\theta), \label{ThetaLD}
\end{equation}
where $\mathbf{\Theta}_0$ is the centroid of the point-images generated by a point-source in $(r \rho_*,\theta)$.

Even in this case, for a constant profile we go back to the centroid of the uniform-brightness disk (\ref{Theta}). Therefore, we can propose to approximate the centroid for the limb-darkened source as the sum of the contributions by a sequence of annuli
\begin{equation}
\mathbf{\Theta}_{LD}\simeq \frac{1}{\mu_{LD}}\sum_{i} \bar{\mathbf{X}}_i, \label{centroidLD}
\end{equation}
with
\begin{equation}
\bar{\mathbf{X}}_i=\frac{F(r_{i})-F(r_{i-1})}{r^{2}_{i}-r^{2}_{i-1}}[\mathbf{X}_i r^{2}_{i}-\mathbf{X}_{i-1}r^{2}_{i-1}]
\end{equation}
and $\mathbf{X}_i$ is the numerator of the centroid for a uniform disk of radius $r_i$, as defined by Eq. (\ref{XIorig}).

As illustrated by \citet{Bozza1}, the number of annuli can be chosen adaptively by suitable error estimators for the limb-darkened magnification. The radii of the annuli are chosen so as to define equal-size intervals in the cumulative function: the new radius $\bar{r}$ between $r_i$ and $r_{i+1}$ is chosen so as to have $F(r_{i+1})-F(\bar{r})=F(\bar{r})-F(r_{i})$. 

In the end, we can easily check the accuracy reached by our centroid calculation (\ref{centroidLD}) with the same strategy described in Section 
 \ref{Sec accuracy}. It turns out that our estimate for the the centroid error (\ref{centroiderror}) remains valid even in the presence of limb darkening.

The best choice of the limb darkening law may vary depending on the photometric accuracy and the source star in a microlensing event. In \texttt{VBBinaryLensing} we have now implemented four popular limb darkening laws \citep{Claret2011}. If we set $\nu=\sqrt{1-r^2}$, we may choose the linear law
\begin{equation}
\frac{I(\nu)}{I(1)}=1-a_1(1-\nu), \label{LDlinear}
\end{equation}
the quadratic law
\begin{equation}
\frac{I(\nu)}{I(1)}=1-a_1(1-\nu)-a_2(1-\nu)^2, 
\end{equation}
the square root law
\begin{equation}
\frac{I(\nu)}{I(1)}=1-a_1(1-\nu)-a_2(1-\sqrt{\nu}),
\end{equation}
the logarithmic law
\begin{equation}
\frac{I(\nu)}{I(1)}=1-a_1(1-\nu)-a_2\ln \nu,
\end{equation}
or even define an arbitrary limb darkening profile.

In Fig. \ref{figlimb}, we show the effects of limb darkening on the centroid trajectory for the geometry described by Fig. \ref{figsource}. In this plot we use the linear limb darkening law (\ref{LDlinear}) and vary the coefficient $a_1$. Indeed, the effects are really tiny and would require extremely accurate astrometry to be appreciated. In realistic microlensing observations, limb darkening would be thus mainly detected and constrained by photometry.

We close this section by noting that Green's theorem can be extended to non-uniform sources and incorporate the description of limb darkening in the choice of function $L_1$ and $L_2$, which become 1-d integrals in the lens plane \citep{Dominik1998}. Such solution avoids the repetition of the full calculation on different annuli, although each individual contribution to the contour integration becomes itself an integral. A comparison between the two strategies would be certainly interesting and could be the goal of a future investigation.

\section{ASTROMETRIC MICROLENSING  FOR A SINGLE LENS}\label{single}

All microlensing calculations for a single-lens and a uniform extended source can be performed in terms of elliptic integrals: both the magnification \citep{Witt1} and the centroid shift  \citep{MaoWitt1998}. However, the numerical implementation of elliptic integrals is relatively demanding and there is a danger of encountering numerical noise in some limits. Therefore, \citet{Bozza2} proposed to use pre-calculated tables that are built using the formulae by \citet{Witt1} for the magnification. More in detail, for a source with radius $\rho_*$ and distance $u$ from the center of the lens, the magnification is

\begin{equation}
\mu(u,\rho_*)=\left\{ 
\begin{array}{ll}
\frac{u^2+2}{u\sqrt{u^2+4}} f_o(\rho_*/u,\rho_*) & u>\rho_* \\
\sqrt{1+\frac{4}{\rho_*^2}} f_i(u\rho_*,\rho_*) & u<\rho_* 
\end{array} 
\right.,
\end{equation}
where the functions $f_o$ and $f_i$ contain the details of the ellpitic integrals by \citet{Witt1} and are tabulated in a separate file. With this approach, any individual evaluations of the magnification are almost immediate.

For the centroid shift we pursue a similar strategy based on the analytical formulae by \citet{MaoWitt1998}. The centroid shift is always in the direction of the source and is given by

\begin{equation}
\Theta(u,\rho_*)=
\left\{ 
\begin{array}{ll}
\frac{u(u^{2}+3)}{u^{2}+2} F_o(\rho_*/u,\rho_*) & u>\rho_* \\
\left(1-\frac{1}{4+\rho_{*}^{2}}\right)u F_i(u\rho_*,\rho_*) & u<\rho_* 
\end{array} 
\right., \label{LDcorrection}
\end{equation}
where the two functions $F_o$ and $F_i$ contain the elliptic integrals and are tabulated separately for fast computation. In Fig. \ref{figESPL} we show the two functions with different colors representing different source radii. 

Finally, for limb-darkened sources, we repeat the calculation on concentric annuli as already explained in Section \ref{Sec limb}.

\section{Conclusions}\label{summary}

The need for new codes dedicated to microlensing has received a great impulse in the last few years. 

The preparation work for the future {\it Roman} mission requires efficient and accurate modeling platforms to face large-scale data flows. Simulations predicting the expected yields in terms of bound planets \citep{Penny2019} and free-floating planets \citep{Johnson2020} strongly rely on the heavy use of such microlensing codes to produce estimates that play a central role in driving the scientific strategy of this space mission. The availability of the same code to make reliable predictions on astrometric shifts and photometric light curves of the microlensed sources will greatly increase the impact of these simulations and help the discrimination of degenerate models.

In parallel, the recent observations of astrometric shifts in microlensing events \citep{Sahu2017,Zurlo2018} have renewed the interest in astrometric microlensing and its possible use for measuring the masses of isolated objects. Finally, the astrometric microlensing events expected from the analysis of Gaia data \citep{McGill2020} will require adequate modeling by a complete code that allows magnification and astrometry calculations at the same time with inclusion of all possible relevant effects.

In this paper we have presented a full implementation for astrometric microlensing fully based on contour integration. We have exploited Green's theorem to express the centroid integrals in terms of line integrals on the boundaries of the images. These are calculated to third order in the arc length by the inclusion of the parabolic correction. The accuracy of the code has been widely tested and is properly tracked by a simple formula that catches its dependencies on the code parameters. Limb darkening is treated by repeating the calculation on concentric annuli and has been extended to arbitrary source profiles. For single-lens microlensing, we have adopted pre-calculated tables that guarantee the maximum speed. As an aside, as shown in the appendix, we have also added  the implementation of full Keplerian orbital motion within \texttt{VBBinaryLensing}, which is a interesting add-on both for photometric and astrometric studies \citep{Sajadian2014}. 

The final goal of our efforts is to put the whole microlensing community in the best possible position to face the challenges set by the new exciting observations that will come at the onset of the new decade. Such detections will greatly enhance our knowledge about extrasolar planets, with the census of planets beyond the snow line, the abundance of free-floating planets and the distribution of planets throughout our Galaxy.

\section*{Acknowledgements}

We thank Weicheng Zang for pre-release testing of the limb darkening code on a real event and Markus Hundertmark for useful feedback. 

\section*{Data Availability}

The code presented in this paper is available as a single zip file at \url{http://www.fisica.unisa.it/GravitationAstrophysics/VBBinaryLensing.htm} and with a full \texttt{Python} implementation at \url{https://github.com/valboz/VBBinaryLensing}.

\bibliographystyle{mnras}
\bibliography{references1}

\appendix
\section{Efficient parameterization for Keplerian orbital motion}\label{kepler}

The orbital motion of the lens must be taken into account whenever the microlensing event has a duration comparable with the orbital period of the planet. In particular, it is typically detected when two distinct features in the caustics are met by the source with a sufficiently long time interval in-between. This is also true for the astrometric centroid discussed in this paper \citep{Sajadian2014}. In general, the shape of the caustics varies with the separation $s$ of the two lenses and their orientation $\alpha$ with respect to the source trajectory. This means that to first order it is possible to catch the main features of the model just by two parameters $ds/dt$ and $d\alpha/dt$. A linear extrapolation based on these parameters does not correspond to any physically acceptable orbits and may lead to unphysical light curves if it is used beyond its range of validity. For this reason, \texttt{VBBinaryLensing} offers orbital motion including the relative radial velocity of the lenses at a reference time as a third parameter. Following the notation of the appendix of \citet{Skowron}, the three components of the velocity vector of the second component relative to the first one at a reference time $t_{0,kep}$ are 
\begin{equation}
\mathbf{v}= R_{E} s (\gamma_{1}, \gamma_{2}, \gamma_{3}),
\end{equation}
with $R_E=D_L\theta_E$ the projected Einstein radius of the system in the lens plane \citep{Paczynski1986}. In our reference frame, the two components in the lens plane introduced in Section \ref{SecMag} are complemented by the third component along the line of sight, pointing away from the observer. The instantaneous relative position at the same reference time is 
\begin{equation}
\mathbf{r}=R_{E} (s,0,s_z)
\end{equation}
We may identify $\gamma_1=\left.\frac{1}{s}\frac{ds}{dt}\right|_{t_{0,kep}}$, $\gamma_2=\left.\frac{d\alpha}{dt}\right|_{t_{0,kep}}$. With three components of the velocity and assuming circular orbital motion, the relative separation along the line of sight is fixed by the condition $\mathbf{r}\cdot \mathbf{v}=0$, which yields $s_z=-s \gamma_1/\gamma_3$, and the orbital motion remains completely specified. The radial velocity is very often left  unconstrained by the data, since it only determines the evolution of the lens at very distant times, when the microlensing signal is generally negligible. However, the assumption of circular orbital motion lets us move within a secure physical framework.

In the new version of the code, we have also included full Keplerian orbital motion, which may be of interest in exceptional situations in which there is enough sensitivity to all orbital parameters. For this case we introduce two additional parameters: the ratio of the radial separation with the projected separation $r_s=s_z/s$ and the ratio of the semimajor axis with the current separation at the reference time $a_s=a/r$. The set of orbital parameters then becomes $(\gamma_1,\gamma_2,\gamma_3,r_s,a_s)$, from which we can derive all orbital elements of the system while remaining in the domain of bound elliptic orbits.

In particular, combining the expression of the mechanical energy (\textit{vis-viva} equation)
\begin{equation}
-\frac{G M}{2a}=\frac{v^{2}}{2}-\frac{G M}{r},
\end{equation}
the Kepler's third law ($n=2\pi/T$ being the orbital angular velocity) 
\begin{equation}
n=\sqrt{\frac{G M}{a^{3}}},
\end{equation}
and our definition of the semimajor axis ratio $a_s$, we find an explicit expression of the orbital velocity as a function of our basic parameters
\begin{equation}
n= \frac{\sqrt{\gamma_{1}^{2}+\gamma_{2}^{2}+\gamma_{3}^{2}}}{a_{s} \sqrt{(-1+2a_{s}) (1+r_s^{2})} },
\end{equation}
and an expression for the total mass as a function of the basic parameters and the Einstein radius $R_E$
\begin{equation}
GM= \frac{R_E^3s^3a_s\sqrt{1+r_s^2}}{2a_{s}-1 }\left(\gamma_{1}^{2}+\gamma_{2}^{2}+\gamma_{3}^{2}\right).
\end{equation}
The mass of the system can be fixed only if we have separate information on the Einstein radius, which may come from the study of the source, astrometric measurements, parallax, lens observation and so on.

In order to derive the orbital parameters, we introduce the specific angular momentum
\begin{equation}
\mathbf{h}=\mathbf{r}\times\mathbf{v},
\end{equation}
and the reduced Laplace-Runge-Lenz vector
\begin{equation}
\mathbf{e}=\frac{\mathbf{v}\times\mathbf{h}}{G M} -\frac{\mathbf{r}}{r}.
\end{equation}
Eliminating the mass from the previous equations, this vector is fully expressed in terms of the basic parameters $\gamma_1,\gamma_2,\gamma_3,r_s,a_s$. Its modulus represents the orbital eccentricity $e=|\mathbf{e}|$, while its direction points toward the periastron. Therefore, we introduce the following orthonormal unit vectors
\begin{equation}
\mathbf{\hat{x}}=\frac{\mathbf{e}}{e}, \quad  \mathbf{\hat{z}}=\frac{\mathbf{h}}{h},  \quad \mathbf{\hat{y}}=\mathbf{\hat{z}}\times \mathbf{\hat{x}},
\end{equation}
which define the orientation of the orbit in space. The three angles $\Omega, i, \omega$ can be read from the components of these vectors: 
\begin{eqnarray}
&&\cos i= z_3 \\
&&\tan \Omega = -\frac{z_1}{z_2}\\
&& \tan \omega = \frac{x_3}{y_3}
\end{eqnarray}

At this point, we can easily obtain the true anomaly $\nu$ by
\begin{equation}
\cos(\nu)=\frac{\mathbf{r}}{r}\cdot \mathbf{\hat{x}}
\end{equation}
and determine its sign from the sign of $\mathbf{r}\cdot \mathbf{\hat{y}}$.
Then we calculate the eccentric anomaly
\begin{equation}
\cos E= \frac{\cos \nu+e}{1+e \cos\nu}
\end{equation}
and the periastron time
\begin{equation}
t_{p}=t_{0,kep}-\frac{E-e \sin E}{n}.
\end{equation}

Once the orbital elements are found, we can easily proceed in the reverse way to find the relative position of the two components at any time, starting from the numerical solution of the Kepler equation
\begin{equation}
E-e \sin E =n (t-t_{p}).
\end{equation}

Our parametrization has several advantages: it makes immediate contact with existing linear approximations, since it includes $\gamma_1=\left.ds/dt/s\right|_{t_{0,kep}}$ and $\gamma_2=\left. d\alpha/dt\right|_{t_{0,kep}}$; circular orbital motion is simply recovered for $r_s=\gamma_1/\gamma_3$ and $a_s=1$; the two ratios $r_s$ and $a_s$ should be numbers of order 1 in most realistic orbits, unless we are observing the system at some special time and orientation. Therefore, they are very convenient for downhill searches and a successful exploration of the parameter space.
\end{document}